\documentstyle[aps,prl,psfig,floats]{revtex}
\newcommand{\lessim}{\mathop{\,<\kern - 1.05 em \lower 1.ex
\hbox {$\sim$}\,}}
\newcommand{\grtsim}{\mathop {\,> \kern - 1.05 em \lower 1.ex
\hbox {$\sim$}\,}}
\begin{document}
\draft
\twocolumn[\hsize\textwidth\columnwidth\hsize
\csname @twocolumnfalse\endcsname
\title{Analysis of the resonance peak
and magnetic coherence \\ seen in inelastic neutron scattering
of cuprate superconductors:\\
a consistent picture with tunneling and conductivity data}
\author{D. Manske$^1$, I. Eremin$^{1,2}$, and K.H. Bennemann$^1$}
\address{$^1$Institut f\"ur Theoretische Physik,
Freie Universit\"at Berlin,
Arnimallee 14, D-14195 Berlin, Germany}
\address{$^2$Max-Planck-Institut f\"ur Physik komplexer Systeme,
N\"othnitzer Str. 38, D-01187 Dresden, Germany}
\date{\today}
\maketitle
\begin{abstract}
Assuming the exchange of antiferromagnetic spin fluctuations
as the Cooper pairing mechanism we calculate the doping
dependence of the resonance peak seen
in inelastic neutron scattering
and the magnetic coherence effect.
Most importantly, we find that the
resonance peak in the magnetic susceptibility,
$\mbox{Im }\chi({\bf q},\omega)$,
appears {\it only} in the superconducting state,
that it scales with $T_c$, and that magnetic coherence is
a result of a $d$-wave order parameter. We further
analyze the structure of $\mbox{Im }\chi$ below $T_c$,
the position of the peak at $\omega_{res}$ and the
consequences for photoemission, tunneling spectroscopy
and the optical conductivity.
\end{abstract}
%
%
]
\narrowtext
For analyzing the pairing mechanism in high-$T_c$
superconductors it is important to understand the
spin-excitation spectrum as observed by
inelastic neutron scattering (INS) \cite{fong,he}.
In particular the doping and temperature dependence of
the spin susceptibility $\mbox{Im}\,\chi({\bf q},\omega)$
and its relationship to the superconducting transition
temperature $T_c$ are important.
INS experiments show the appearance of a resonance peak
at $\omega_{res}$ only below $T_c$ \cite{fong} and find a
constant ratio of $\omega_{res}/T_c \simeq 5.4$ for
underdoped $\mbox{YBa}_2\mbox{Cu}_3\mbox{O}_{7-\delta}$
(YBCO) and overdoped $\mbox{Bi}_2\mbox{Sr}_2
\mbox{CaCu}_2\mbox{O}_{8+\delta}$ (BSCCO)
\cite{he,remark1,dai,keimer}. Recent INS data on
$\mbox{La}_{2-x}\mbox{Sr}_x\mbox{CuO}_4$
(LSCO) reveal strong
momentum- and frequency-dependent changes of
$\mbox{Im}\,\chi({\bf q},\omega)$ in the
superconducting state which the authors called
magnetic coherence effect \cite{masonlake}.
In particular, $\mbox{Im}\,\chi ({\bf Q_i})$
with ${\bf Q}_i= (1 \pm \delta,1 \pm \delta)\pi$
is strongly suppressed compared to its normal
state value below $\omega\lessim 8$ meV, while it increases
above this frequency. Furthermore, the incommensurate peaks
become sharper in the superconducting state \cite{masonlake}.
Several theoretical approaches have been considered for
the resonance peak \cite{bulut,demler,levin,manske2,lee},
its relation to the spectral function seen in ARPES and
tunneling data \cite{norman1,chubukov1},
and for the magnetic coherence \cite{morr}, but no unified
theory has been presented.
Here, using an electronic theory, we perform calculations
to demonstrate the significance of the feedback of
superconductivity on $\mbox{Im }\chi$. This
permits us to derive for underdoped and overdoped cuprates
the relationship between different
measurements like optical conductivity and tunneling.
So far the resonance peak and magnetic coherence have 
been treated separately and in both cases without
taking into account this
important  feedback of superconductivity
on $\mbox{Im }\chi$. 

In this Communication we use an 
electronic theory for the spin susceptibility and
for the Cooper pairing via exchange of
antiferromagnetic spin fluctuations
to analyze the consequences of the superconducting feedback
on magnetic coherence and the resonance peak, and on the
relationship between INS, tunneling, and optical conductivity. 
The investigation of the effect of superconductivity on
$\mbox{Im }\chi$ is an extension of our
previous work \cite{manske2,langer}.
Using RPA and self-consistent FLEX
\cite{bickerswhite} calculations for
$\mbox{Im}\,\chi({\bf q},\omega)$, we present results for the
kinematic gap (or spin gap) $\omega_0$,
$\omega_{res}$, $\omega_{res}/T_c$,
and the gap function $\Delta(\omega)$ in reasonable
agreement with experiments.
Most importantly, we find that our electronic theory can explain
consistently within the same picture inelastic neutron scattering,
optical conductivity, and SIN tunneling data.
Furthermore, the same physical picture 
gives results for underdoped and
overdoped cuprates \cite{schachinger,manske}.

In order to analyze the kinematic gap and the position of
the resonance peak it is instructive to start
with the bare Lindhard BCS susceptibility
\cite{bulut} at ${\bf q}={\bf Q}=(\pi,\pi)$
%
%
\begin{eqnarray}
\mbox{Im}\,\chi_0({\bf Q}, \omega)
& = &
\frac{1}{2} \sum_k \left\{ (1-2f(E_k))
\, \delta (\omega + 2E_k) \right. \nonumber\\
& + &
\left.
( 2f(E_k) - 1 ) \, \delta (\omega - 2E_k) \right\}
\label{eq:chi0}
\quad ,
\end{eqnarray}
%
%
where $f(E_k)$ is the Fermi function and
$E_{\bf k}=\sqrt{\epsilon_{\bf k}^2+\Delta_{\bf k}^2}$ is the
dispersion of the Cooper-pairs in the superconducting
state. We use a gap function with $d$-wave symmetry,
$\Delta_{\bf k}= \Delta_0 (\cos k_x - \cos k_y) / 2$,
which can be calculated self-consistently within our
FLEX-approach.
For the normal state dispersion, we employ a tight-binding
band
\begin{equation}
\epsilon_{\bf k}=-2t\left[\cos k_x + \cos  k_y
-2t'\cos k_x \cos k_y - \mu/2 \right]
.
\label{eq:eps}
\end{equation}
Here, $t$ is the nearest neighbor hopping energy, $t'$ denotes
the ratio of next-nearest neighbor to nearest neighbor hopping
energy, and $\mu$ is the chemical potential.
We use $t'$ as a fitting parameter in order to describe the
Fermi surface topology of both materials YBCO and LSCO.
In Eq. (\ref{eq:chi0}) we take $t'=0$ and,
for simplicity, we do not consider a bilayer coupling
via a hopping integral $t_\bot$ \cite{remark2,manske3}.
$\mbox{Im}\,\chi_0 ({\bf Q},\omega)$ involves two characteristic
frequencies. The first, $\omega_{DOS}$,
arises from the density of states
of the Bogoluibov quasiparticles
(i.e. the Cooper-pairs) which have a gap in their
\begin{figure}[ttt]
\vspace{-1.0cm}
\centerline{\psfig{file=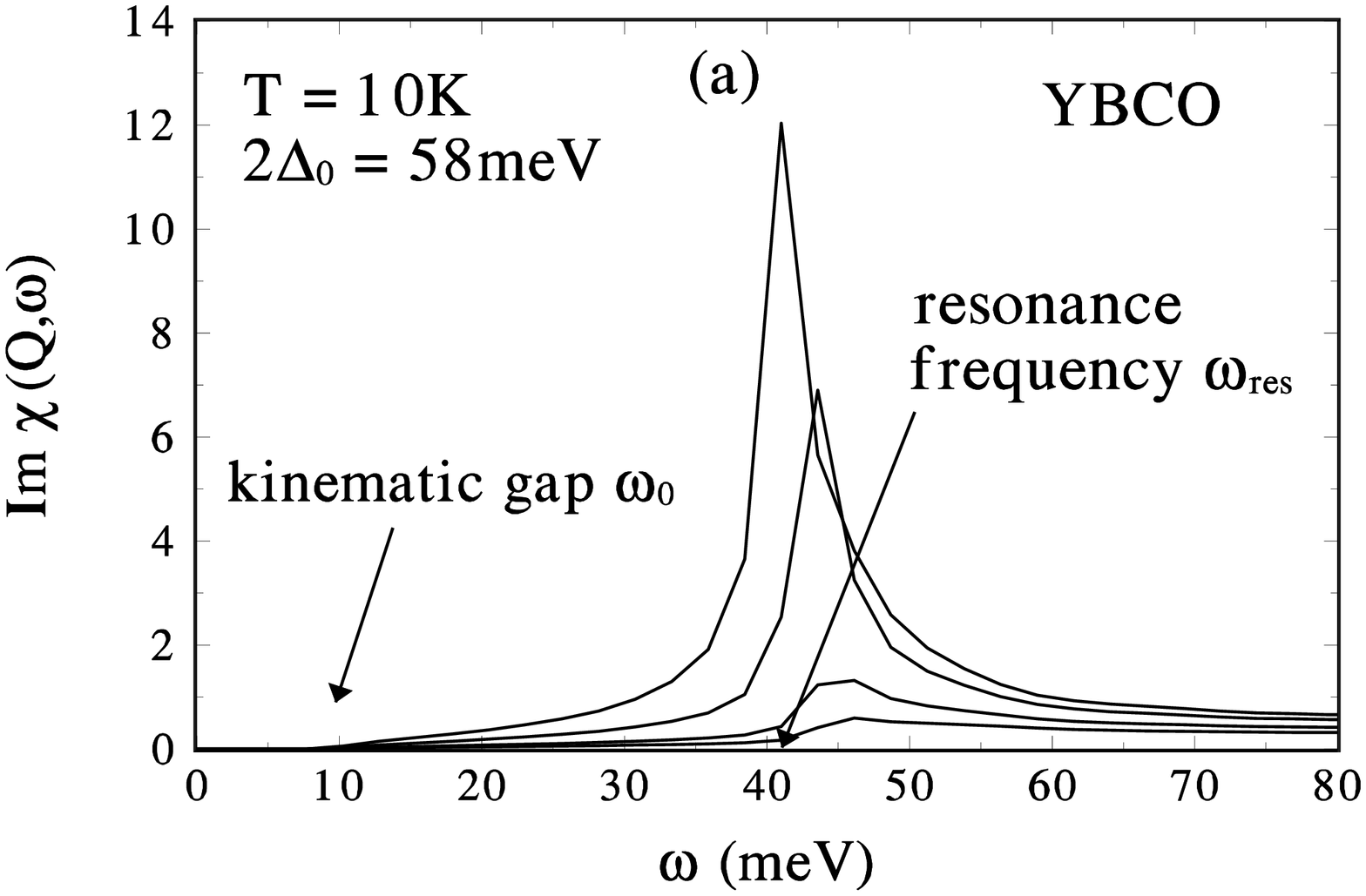,width=8.5cm,angle=-90}}
\vspace{-0.2cm}
\centerline{\psfig{file=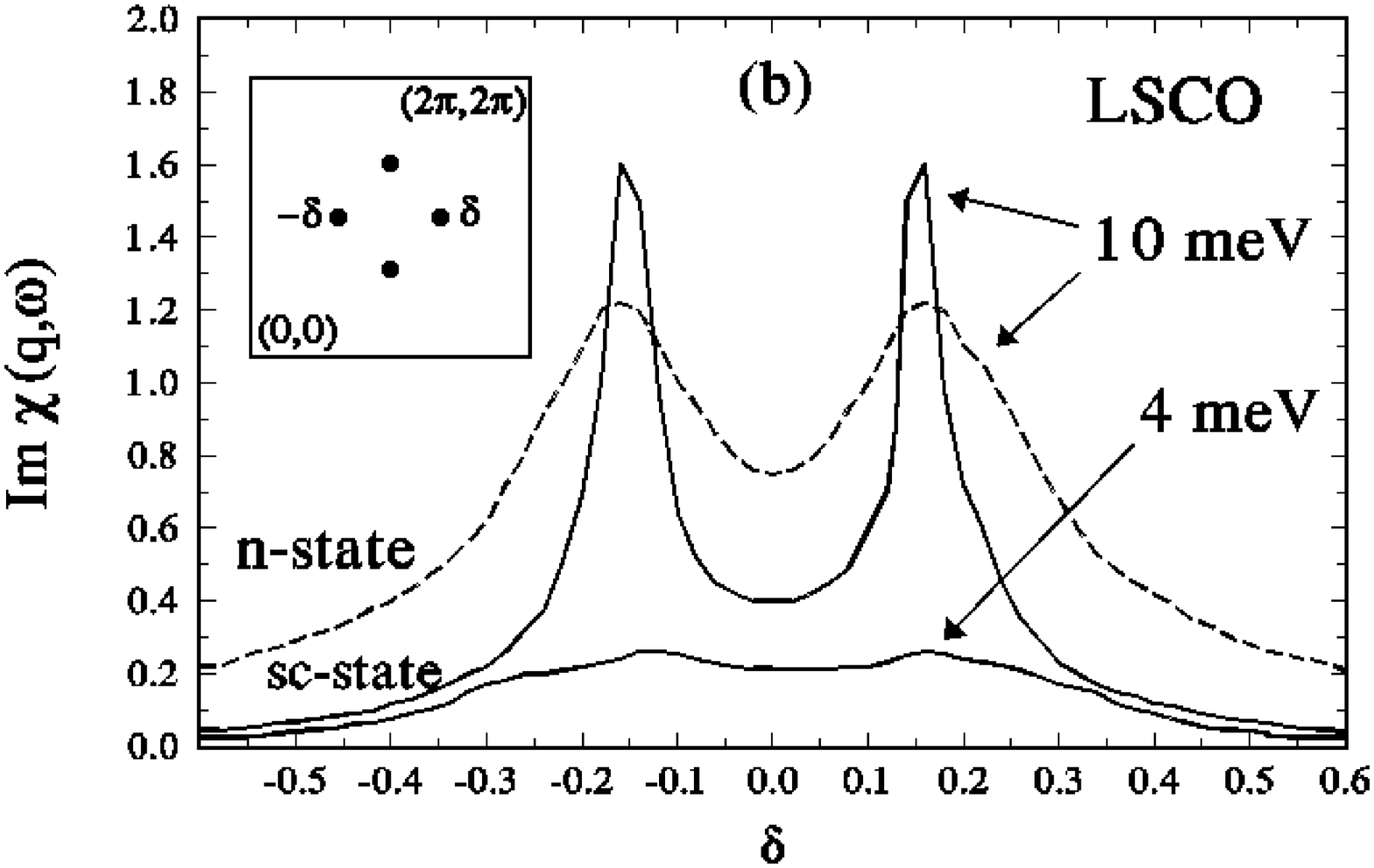,width=8.5cm,angle=-90}}
\vspace{1ex}
\caption{Numerical results for
the resonance peak and magnetic
coherence in the weak-coupling limit.
(a) Imaginary part of the RPA spin susceptibility
(in units of states/eV) versus $\omega$ in
the superconducting state
at wavevector ${\bf q}={\bf Q}=(\pi,\pi)$ for
$U/t=1$, $2$, $3$, and $4$ (from bottom to top).
As in Ref.{\protect\onlinecite{fong}}
we find $\omega_{res}=41\mbox{ meV}$.
Below the kinematic gap $\omega_0$,
$\mbox{Im }\chi({\bf Q},\omega)$ is zero.
(b) Calculated magnetic coherence: the solid curves correspond
to the superconducting state whereas the dotted curve is
calculated in the normal state.
The observed four peaks occur at
${\bf Q}_i= (1 \pm \delta,1 \pm \delta)\pi$ and in the figure
we show only the peaks at ${\bf Q}_i= (1,1 \pm \delta)\pi$.
In our calculations we find $\delta \approx 0.18$.
These results are in fair agreement with experiments,
see Ref. \protect\onlinecite{masonlake}.}
\label{fig1}
\end{figure}
spectrum due to superconductivity, $\omega_{DOS}\simeq
2\Delta(x,T)$. Here, $x$ is the doping concentration.
The second, $\omega_0$, at which
$\mbox{Im}\,\chi_0 ({\bf Q},\omega)$ starts to increase represents the
existence of a $d$-wave superconducting order parameter and is
the so-called kinematic gap \cite{bulut,manske3}.
We point out that, for $t'>0.3$ the kinematic gap is washed out.

In Fig. \ref{fig1} (a) results for the spin susceptibility
%
\begin{eqnarray}
\lefteqn{
\mbox{Im}\,\chi({\bf Q},\omega)
= }
\nonumber\\
&  &
\frac{\mbox{Im}\,\chi_0 ({\bf Q},\omega)}
 {(1-U\,\mbox{Re}\,\chi_0 ({\bf Q},\omega))^2
 +U^2\,(\mbox{Im}\,\chi_0({\bf Q},\omega)^2}
\quad ,
\label{chirpa}
\end{eqnarray}
%
%
are shown . Here,
$U$ stands for an effective Hubbard interaction.
We choose $t'=0.2$ in order to describe the Fermi surface
topology of both YBCO and LSCO.
Generally, one finds that the structure of $\mbox{Im }\chi$
is determined by $\mbox{Im }\chi_0$ if $(U\mbox{Re }\chi_0)\neq 1$
and by $(U\mbox{Re }\chi_0) = 1$ if this can be fulfilled.
We again find the two characteristic frequencies
$\omega_0$ and $\omega_{res}\simeq\omega_{DOS}$ at which
$\mbox{Im}\,\chi$ is peaked.
Furthermore, one clearly sees that
for increasing $U$ the peak in $\mbox{Im }\chi$ shifts to lower
energies, and most importantly, becomes resonant for $U=U_{cr}$
which satisfies the condition
\begin{equation}
\frac{1}{U_{cr}} = \mbox{Re }\chi_0({\bf q}={\bf Q},\omega=
\omega_{res})
\quad ,
\label{ucr}
\end{equation}
which signals the occurence of a spin-density-wave collective
mode. The real part 
is given (at $T=0$) by:
\begin{eqnarray}
\lefteqn{
\mbox{Re }\chi_0({\bf Q},\omega_{res}) = }
\nonumber\\
&  &
\sum_{\bf k}
\frac{E_{\bf k}E_{{\bf k}+{\bf Q}} -
\epsilon_{\bf k}\epsilon_{{\bf k}+{\bf Q}} -
\Delta_{\bf k}\Delta_{{\bf k}+{\bf Q}}}{\left(
E_k + E_{k+Q}\right)^2 - \omega^2}\,
\frac{E_{\bf k} + E_{{\bf k}+{\bf Q}}}{2E_{\bf k}
E_{{\bf k}+{\bf Q}}}
\quad .
\label{rechi}
\end{eqnarray}
\begin{figure}[t]
\vspace{-1.0cm}
\centerline{\psfig{file=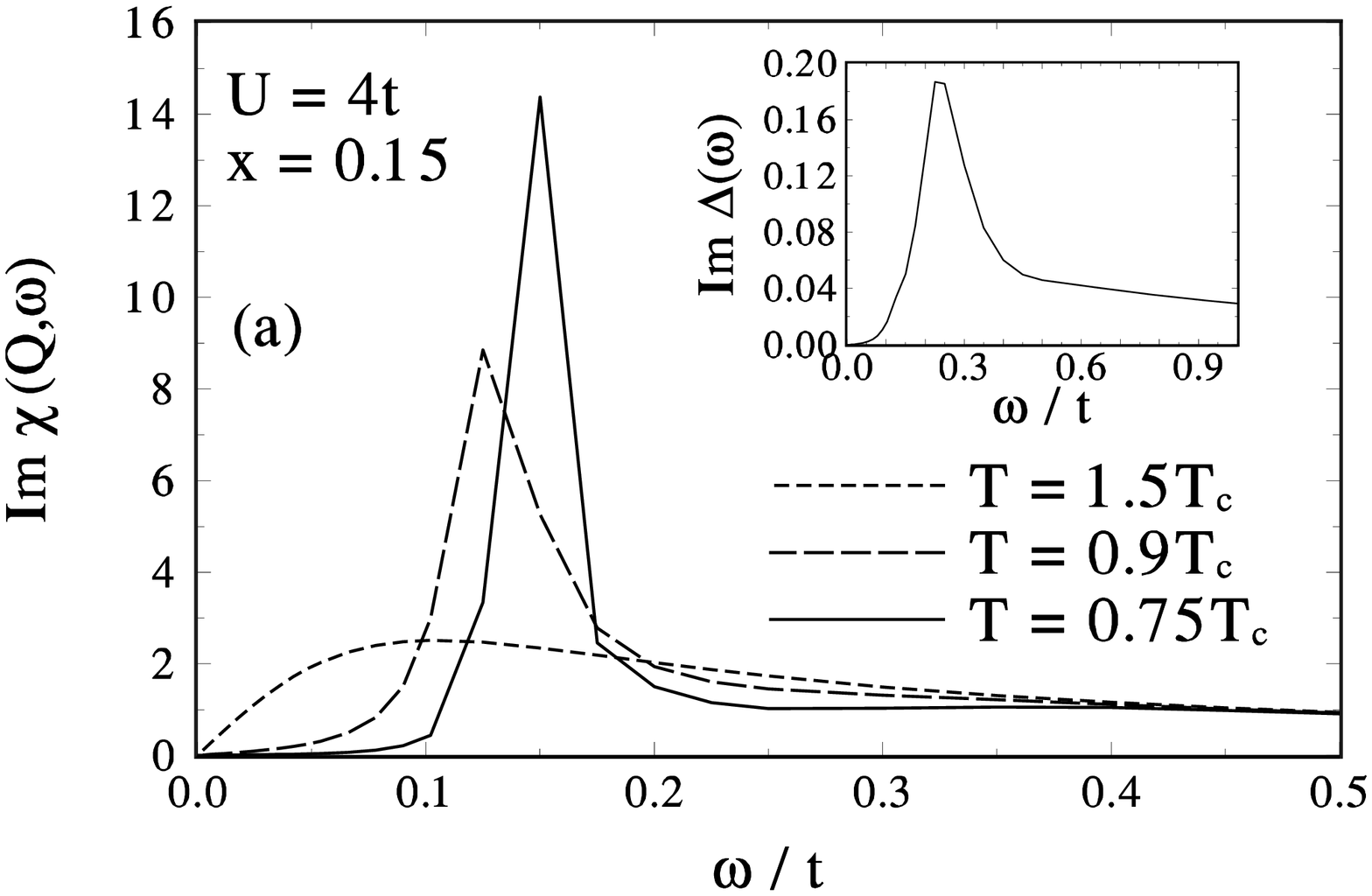,width=8.5cm,angle=-90}}
\vspace{-1.0cm}
\centerline{\psfig{file=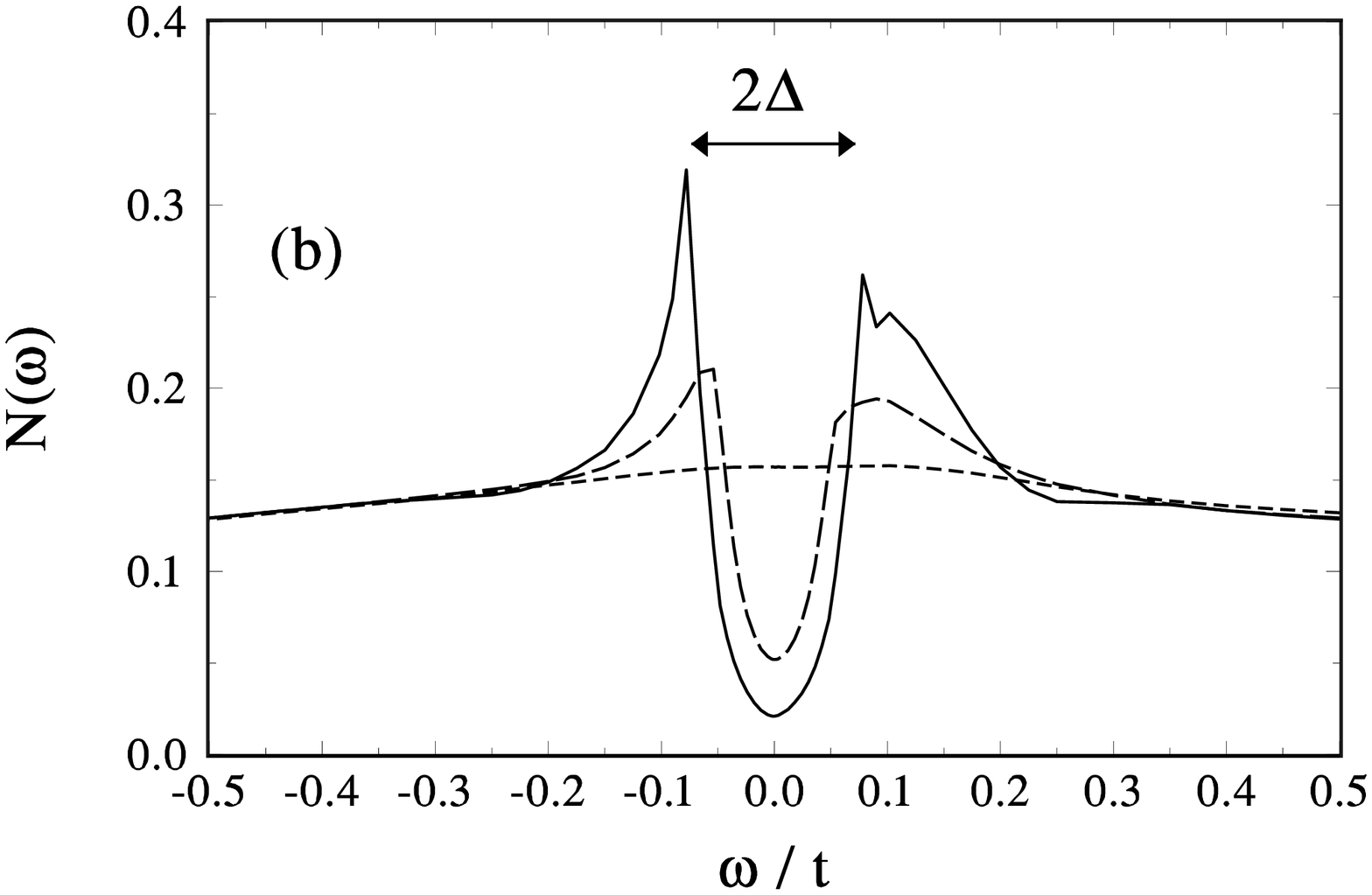,width=8.5cm,angle=-90}}
\vspace{1ex}
\caption{Consistent picture of INS and tunneling data.
(a) Imaginary part of the RPA spin susceptibility
at ${\bf q}={\bf Q}=(\pi,\pi)$
calculated within the FLEX approximation.
For the normal state (dotted) we get
$\omega_{sf} = 0.1t$ and in the superconducting state
we obtain $\omega_{res}=0.15t$.
Assuming $t=250\mbox{ meV}$ we find that $0.16t=40\mbox{ meV}$.
Inset: Imaginary part of the gap function at $T=0.075T$ for
wave vector ${\bf q} \simeq (\pi,0)$.
(b) Calculated density of states for the same parameters and
temperatures as in (a).}
\label{fig2}
\end{figure}
$\mbox{Re }\chi_0({\bf Q},\omega_{res})$
has been investigated in detail in Ref.
\onlinecite{manske2}, where it was found that the
spin-density-wave collective mode, that satisfies Eq. (\ref{ucr}),
can explain the dip and hump feature observed in the photoemission
spectra on BSCCO \cite{norman}. In particular, it was shown
that the broad humps are at the same position for both the
normal and superconducting state.

We find from Eq. (\ref{chirpa}) that,
in the normal state where no
resonance appears, the spin wave spectrum is mainly determined
by the spin fluctuation frequency $\omega_{sf}$
(roughly the peak position) and
\begin{equation}
\mbox{Im}\,\chi({\bf Q},\omega) = \frac{\omega\,\omega_{sf}}{
\omega^2 + \omega_{sf}^2}
\quad .
\label{chioz}
\end{equation}
In the superconducting state one finds that
$\mbox{Im }\chi$ peaks resonantly at $\omega_{res}$
where $\omega_{res}\simeq 2\Delta$ as it
can be already seen from Eq. (\ref{eq:chi0}).
More precisely, we find for {\it optimal doping},
where Eq. (\ref{ucr}) determines the structure
of $\mbox{Im }\chi$ the important relation
%
%
$\omega_{res}(T) = 2\Delta_0(T) - \omega_{sf}(T)$.
Physically speaking, the resonance peak
appearing in INS only below
$T_c$ is mainly determined by the maximum of the superconducting
gap, but renormalized by normal state spin excitations.
\begin{figure}[t]
\vspace{-1.0cm}
\centerline{\psfig{file=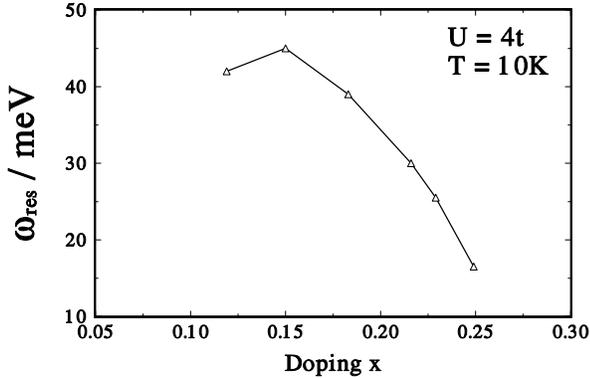,width=8.5cm,angle=-90}}
\vspace{1ex}
\caption{Calculated FLEX results $\omega_{res}$ versus doping.
In the overdoped regime, with
$T_c \propto \Delta_0$ \protect\cite{manske},
we find a constant ratio of
$\omega_{res}/T_c\simeq 8$.}
\label{fig3}
\end{figure}
This provides a simple explanation for the observed $41$ meV
resonance peak in optimally doped YBCO
\cite{fong}, because Raman data
suggest $2\Delta=58$ meV \cite{chen}
and $\omega_{sf}\simeq 17$ meV (at 100 K) as extracted from
NMR experiments \cite{pines}.

In Fig. \ref{fig1}(b) we show results for
the ${\bf q}$-dependence of
$\mbox{Im}\,\chi({\bf q},\omega)$.
We perform our calculations for $U=2t$ and a
superconducting gap of $2\Delta=10\mbox{ meV}$
as measured in Raman scattering in optimally doped
$\mbox{La}_{1.85}\mbox{Sr}_{0.15}\mbox{CuO}_4$ \cite{chen2}.
For $\omega=10$ meV we get two peaks at ${\bf q}={\bf Q}_i$.
In the superconducting state we find a sharpening of the peaks
due to the occurrence of a gap. This simply means that the
lifetime of the quasiparticles is enhanced due to a reduced
scattering rate. At $4$ meV these peaks are strongly
suppressed as seen in the experiment \cite{masonlake}.
Furthermore, we find no signal for $\omega < 4$ meV.
This is due to the kinematic
gap seen in  Fig. \ref{fig1}(a) which is
independent of ${\bf q}$.
Note, the situation would be totally different if LSCO would
have an isotropic gap where all states for
$0 < \omega < 2\Delta_0\simeq 20 \mbox{ meV}$
would be forbidden. In this case
no kinematic gap (or spin gap) would be observed.
Thus, we can conclude from Fig. \ref{fig1}
that, already in the weak coupling limit where
no lifetime of the Cooper-pairs
(i.e., $\Delta$ independent of $\omega$) is
considered, we are able to explain the resonance peak and
the magnetic coherence effect.

In order to consider the important feedback effect
of $\Delta$ on the spin excitation spectrum,
we now discuss our results obtained in the strong-coupling limit
(i.e. $\omega$-dependent) solving self-consistently the
generalized Eliashberg equations
within the FLEX approximation \cite{langer,bickerswhite}.
Note that, only $U/t$ and the tight-binding
dispersion relation  $\epsilon({\bf k})$
(with its band filling $\mu$) enter
the theory as free parameters.

In Fig. \ref{fig2}(a) we present results for
$\mbox{Im}\,\chi({\bf Q},\omega)$ calculated for
$U=4t$ and for an optimum doping concentration
$x=0.15$ ($\mu=1.65$).
In the normal state (dotted curve) we find $\omega_{sf}=0.1t$,
whereas for $T < T_c$
the resonance peak (solid curve) appears at
$\omega_{res}=0.15t$. The dashed curve corresponds to $T=0.9T_c$
where the superconducting gap starts to open. Thus, the peak
position reveals information on the temperature dependence of
the superconducting gap.
For temperatures $T < 0.75T_c$ the resonance peak
remains at $\omega_{res} = 0.15t$ and only the peak
height increases further.
\begin{figure}[t]
\vspace{-1.0cm}
\centerline{\psfig{file=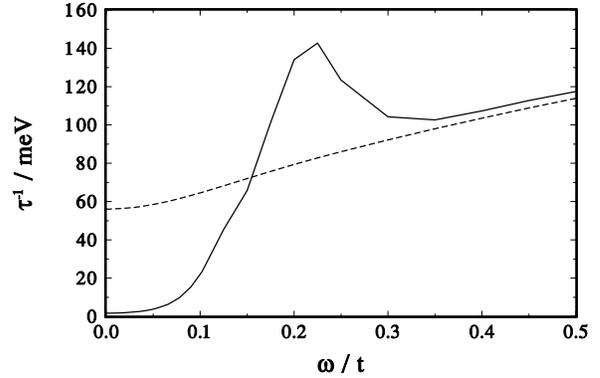,width=8.5cm,angle=-90}}
\vspace{1ex}
\caption{Calculated scattering rate
in the normal state for $T=1.5T_c$ (dotted curve) using
equation (\protect\ref{eq:rho})
and superconducting state $0.75T_c$ (solid curve) using
Kubo formula and yielding a threshold ($2\Delta+\omega_{res}$). 
The results are in fair
agreement with Ref. \protect\onlinecite{timusk}.}
\label{fig4}
\end{figure}
We find that the height of the peak is of the order of the
quasiparticle lifetime $1/\Gamma(\omega_{res})$,
where $\Gamma({\bf k},\omega)=\omega\,\mbox{Im }Z({\bf k},\omega)
/ \mbox{Re }Z({\bf k},\omega)$. $Z$ denotes the mass
renormalization within the Eliashberg theory.
Thus we can conclude that the resonance peak becomes
observable because the scattering rate decreases drastically
below $T_c$ \cite{manske2}.

In Fig. \ref{fig2}(b) we show the corresponding calculated density of
states $N(\omega)$.  Below $T < 0.75T_c$ we find that the value of
$2\Delta$ determined from peak-to-peak stays approximately constant
and is very close to the value $\omega_{res}$ seen in INS, i.e. $41$
meV, as shown in (a).  This is in good agreement with measured SIN
tunneling data in Ref. \onlinecite{fischer}.  However, in SIN
tunneling a renormalized value of $2\Delta$ is observed. Note, a
direct measurement of $\Delta(\omega)$ (e.g. SIS tunneling) would lead
to higher values. For example, we show in the inset of Fig. \ref{fig2}
(a) the imaginary part of gap function at wave vector
${\bf q} \simeq (\pi,0)$ where the gap has its maximum.
It is peaked at $\omega = 0.25t$.

In Fig. \ref{fig3} we show results for $\omega_{res}$ as a
function of the doping concentration.
We find that, for a fixed $U$ Eq. (\ref{ucr})
cannot be fulfilled in the overdoped case \cite{remark4}.
Thus we find that in this regime the resonance peak is
mainly determined by $\mbox{Im }\chi_0({\bf Q},\omega)$
and thus by $2\Delta_0$.
On general grounds one expects $T_c \propto \Delta_0$
in the overdoped regime, where the system behaves mean-field
(BCS) like. Recently, this has been confirmed within the
FLEX approach \cite{manske}. Thus, we conclude that
$\omega_{res}/T_c$ should be a constant ratio.
We find $\omega_{res}/T_c\simeq 8$ which is larger than
the observed value in BSCCO \cite{he}.
This is due to an underestimation of $T_c$ within FLEX.

In contrast to the overdoped case we find in the underdoped
regime, where $T_c \propto n_s$
($n_s$ denotes the superfluid density) \cite{manske},
that the resonance condition Eq. (\ref{ucr}) yields
$\omega_{res} \propto \omega_{sf}$ which is decreasing.
Note that, the superconducting gap guarantees that
Eq.(\ref{ucr}) is fulfilled. Thus we find a decreasing
resonance frequency for decreasing doping in
agreement with earlier calculations \cite{lee,li}.
To summarize our discussion we have the following result:
\begin{equation}
\omega_{res} \approx \left\{
\begin{array} {ll}
\omega_{sf}, & \mbox{underdoped}\\
2\Delta_{0} - \omega_{sf}, & \mbox{optimal doping}\\
2\Delta_{0}, & \mbox{overdoped}
\end{array}
\right.
\label{wres}
\end{equation}
where the optimally doped case corresponds to $x_{opt} = 0.15$
holes per copper site. This predicted doping dependence of the
resonance peak position should be further tested experimentally.

In order to discuss the consequences of our analysis in
particular the  feedback of superconductivity on
$\mbox{Im }\chi$ for various superconducting 
properties we derive \cite{remark3}
\small
\begin{eqnarray}
\mbox{Im }\Sigma(\omega)
& = &
-\frac{U^2}{4}\,\int_{-\infty}^{\infty} d\omega'
\left[\coth\left(\frac{\omega'}{2T}\right) - \tanh\left(
\frac{\omega' - \omega}{2T}\right)\right]
\nonumber\\
& \times &
\mbox{Im }\chi({\bf Q},\omega') \,
\sum_{\bf k}\delta(| \omega - \omega' | - E_{\bf k})
\quad ,
\label{eq:rho}
\end{eqnarray}
\normalsize
where
$N(\omega)=\sum_{\bf k}\delta(|\omega| - E_{\bf k})$
is the density of states.
Eq. (\ref{eq:rho}) is valid in {\it both} the normal and
superconducting state.
It permits discussion of how much $\Sigma(\omega)$ reflects
$\omega_{sf}$ and $\omega_{res}$, for example.
We see that the feedback of superconductivity on $\mbox{Im }\chi$
causes approximately a shift for the elementary excitations
$\omega\rightarrow\omega+\Delta_{0}$
for the superconducting state in 
the spectral density $U^2\,\mbox{Im }\chi({\bf Q},\omega)/4$.
Using Eq. (\ref{chioz})
in Eq. (\ref{eq:rho}) would not take into account the important
feedback of superconductivity on $\mbox{Im }\chi$.
Eq. (\ref{eq:rho}) can be used to demonstrate the
relationship  between INS and optical conductivity
measurements for the normal state, but note for the
superconducting state we calculate $\sigma$ from $(GG+FF)$. 
Using Drude theory we find that the scattering
rate $\tau^{-1}(\omega)$ agrees qualitatively with
$-2\,\mbox{Im }\Sigma(\omega)$ for the normal state. 
However, in order to get a
quantitative agreement with experimental data one has to
use $\tau^{-1}(\omega) = \Gamma({\bf Q},\omega)$.
This is shown in Fig. \ref{fig4}, in a good agreement
with Ref. \onlinecite{timusk}.
From this analysis we can conclude that optical
conductivity data and thus the lifetime of the quasiparticles,
as well as the resonance peak and SIN tunneling data,
can be understood within our electronic theory.

In summary, we are able to explain consistently all
characteristic facts about the spin excitation spectrum and
its doping dependence in
high-$T_c$ cuprates within an electronic theory.
In particular we find that the resonance peak is a rearrangement
of spectral weight of the normal state which happens only below
$T_c$. 
%
%
Furthermore we show that magnetic coherence
is connected with the resonance peak and can be explained by
a kinematic gap and by a $d$-wave symmetry of
the superconducting order parameter.
By taking into account the feedback of
superconductivity on $\mbox{Im}\,\chi({\bf q},\omega)$
we argue that ARPES results,
tunneling data, and measurements of the
optical conductivity are consistent.
This further strengthens spin fluctuation exchange as the
relevant mechanism for Cooper-pairing in high-$T_c$
cuprates.

It is a pleasure to thank A. Chubukov, J. Schmalian, 
E. Schachinger, T. Dahm, D. Fay, and C. Joas for 
stimulating discussions.


\begin{references}
\bibitem{remark1} A closer inspection of the normal-state data
of underdoped $\mbox{YBa}_2\mbox{Cu}_3\mbox{O}_{6+x}$
\protect\cite{dai} shows that this peak is
qualitative different from the resonance peak
\protect\onlinecite{keimer}.
\bibitem{remark2}
Bilayer coupling leads to better nesting of the
bonding and antibonding bands \protect\cite{bulut}.
Thus a resonance peak might
appear also for a complicated band structure.
\bibitem{remark4} Note, away from optimal doping
determining formally the minimum of $1-U\mbox{Re }\chi_0$
one get $\omega_{min}=2\Delta_0-\omega_{sf}$.
However, the physically relevant condition $1=U\mbox{Re }\chi_0$
yields $\omega_{res}$.
\end{references}
\end{document}